# RETGEM with polyvinylchloride (PVC) electrodes

*V.I.Razin, B.M.Ovchinnikov, A.I.Reshetin, S.N.Filippov*

*INR RAS, Moscow*

**Abstract**

This paper presents a new design of the RETGEM (Resistive Electrode Thick GEM) based on electrodes made of a polyvinylchloride material (PVC).

Our device can operate with gains of $10^5$ as a conventional TGEM at low counting rates and as RPC in the case of high counting rates without of the transit to the violent sparks. The distinct feature of present RETGEM is the absent of the metal coating and lithographic technology for manufacturing of the protective dielectric rms.

The electrodes from PVC permit to do the holes by a simple drilling machine.

Detectors on a RETGEM basis could be useful in many fields of an application requiring a more cheap manufacturing and safe operation, for example, in a large neutrino experiments, in TPC, RICH systems and etc.

## 1. Introduction

The main advantages of RETGEM in comparison with a usual TGEM are appeared by a full safe from the sparks [1]. In our case the discharge current is also limited by a high resistivity of the electrodes manufactured from PVC like a captone or RPC glass with a thick layer of the graphite paint. RETGEM can operate with gains of about $10^5$. At a higher gain it may transit to the strimer mode but a discharge current do not exceed the value of 15 mkA without of the damage either the detector or the front-end electronics. At low counting rate RETGEM can operate as a conventional TGEM and as RPC in the case of high counting rates. We have not undertaken the special efforts as to graphite coating. We did not use the additional lithographic technology and the coating by Cu for making of the holes.

The aim of our activity was to build and test the first prototype of RETGEM without the metal electrodes with a simple layer resistance material like PVC with a thick layer of graphite paint. It permits to drill the holes without micro-particles remaining after drilling.

## 2. Experimental setup

A schematic drawing of the experimental setup for testing of RETGEM detectors is shown in Fig.1.

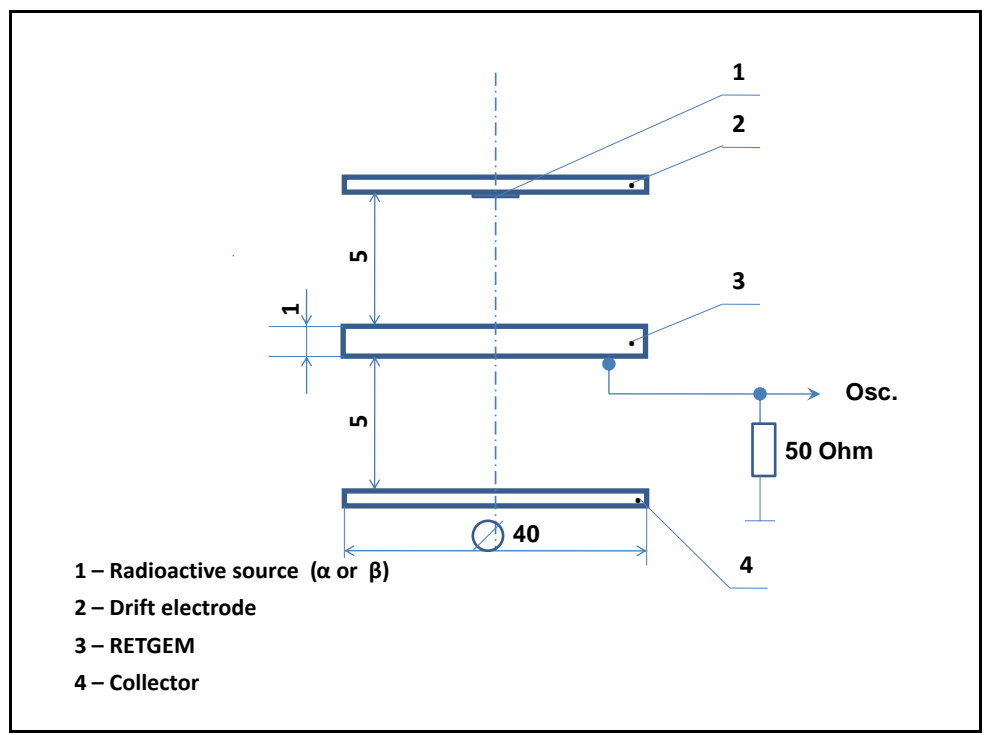

**Fig.1.** A schematic drawing of the experimental setup for testing of the RETGEM

For manufacturing of the electrodes the standard polyvinylchloride of 1mm thickness was used. This material is widely employed in the mass production of the strimer tubes. The surface resistivity of this material may vary from 50 to 1000 k$\Omega$/cm$^2$ by means of a graphite coating paint of 5 – 10 mkm thickness providing a full spark protection. The holes were drilled by usual machine as it was for the case with TGEM in our workshop without of any metallization and etching of the area around holes. The holes were 1.0 mm in diameter with a 2 mm pitch.

The operation principle of this detector was the same as for a case with TGEM. When an HV is applied to the graphite electrodes despite of their non-infinite resistivity they act as equipotential layers and the electrostatic field is formed inside and outside of the holes as in the case of the TGEM with metallic electrodes.

The distances between functional elements of the detector are pointed in Fig.1.

Most of the tests were performed by using Ar + 20%$CO_2$ or Ar + 20% $CH_4$ gas mixture at a pressure of 1 bar. The gas ionization was produced either by a α-source or β-source. The induced signals from the collector were measured by a oscilloscope without any amplifier. By means of this method we prevented the detector from sparks and deterioration of his structure.

## 3. Results

Fig.2 and 3 show the amplitude-voltage dependence measured in the single RETGEM operating in Ar + 20%$CO_2$.

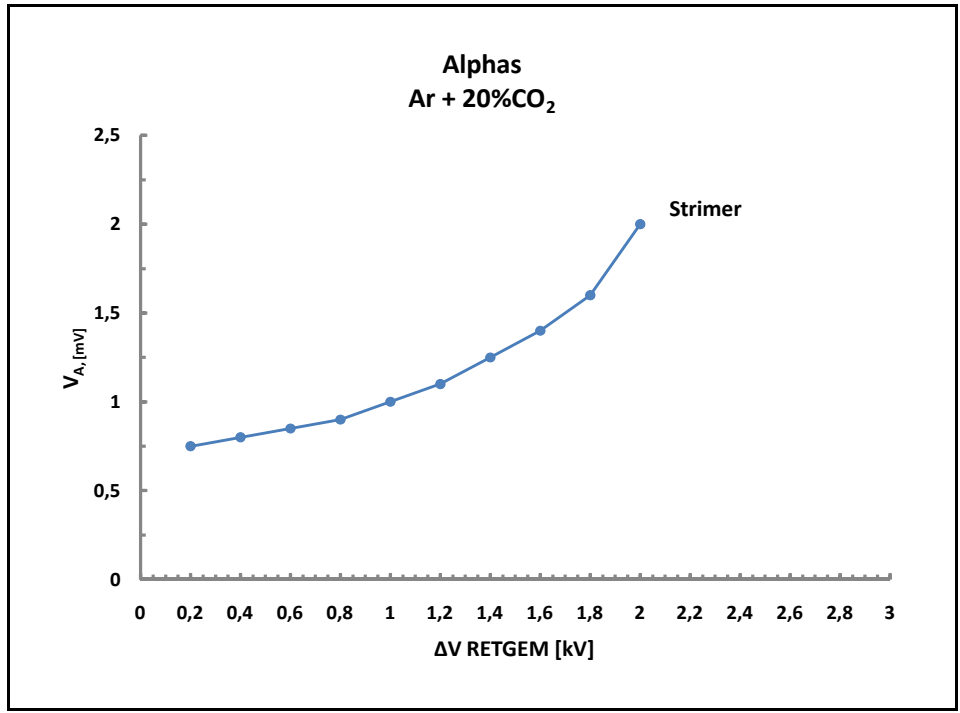

**Fig.2.** The amplitude of the signal v.s. voltage with the RETGEM operating in Ar + 20%$CO_2$ using α-source with I = $10^4$ particles/s.

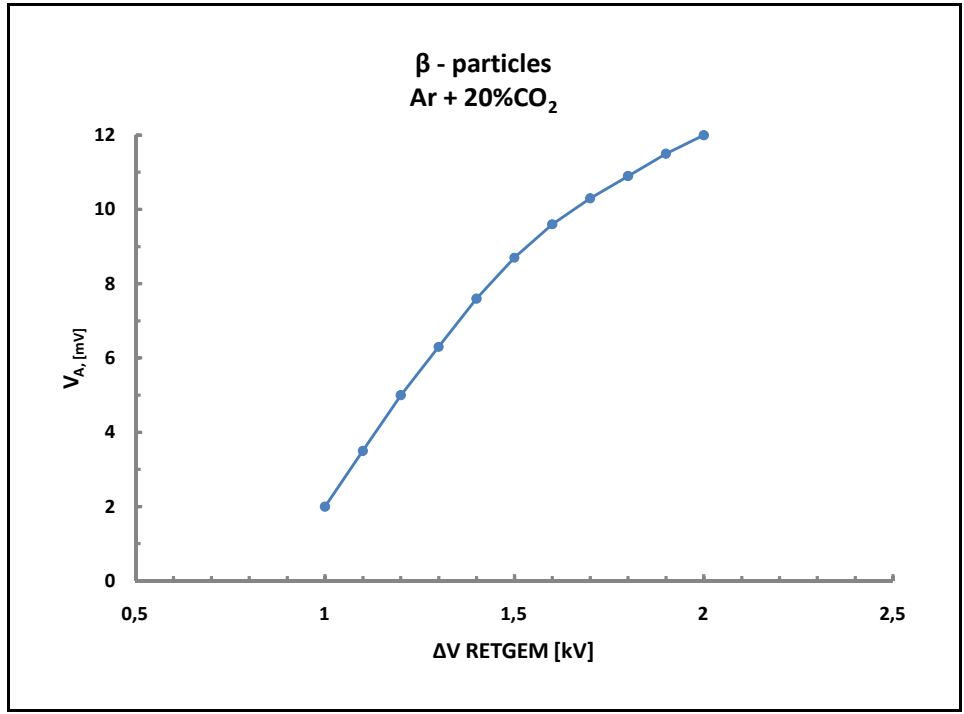

**Fig.3.** The amplitude of the signal v.s. voltage as a Fig.2 using β-source with I = $10^3$ particles/s.

The signal amplitude for the case of the α-source with I = $10^4$ particles/s is not large. It can be explained by reaching of the Rather limit:

$$n_0 \times A \geq 10^8,$$

where $n_0$ – number of initial electrons and A – amplification factor.

Under these conditions we have limited a proportional mode or strimer mode. The current measurements showed that the current of the RETGEM was not more than 15 – 20 mkA. Thus, the sparks current was eliminated in our device.

In the case of the β-source the gain as a function of the difference voltage applied to the RETGEM electrode close to $10^5$ (it is corresponding to the signal amplitude of about 10 mV in our measurements) was achieved.

It is interesting to note, that there was a kind of saturation in the curve merging to gain value of about $10^6$, but the breakdown was not appeared.

## 4. Conclusions

It was observed in this work that a single-layer RETGEM with electrodes made of PVC was fully spark-protected. The achieved gain of $10^5$ is sufficient for more applications that require safe and high operations.

We confirmed that RETGEM is very robust device. It does not require of special cleanliness of its surfaces and can operate with poorly quenched gases.

From point of view the widely using of PVC in mass production of the strimer tubes our proposed RETGEM also has a good perspectives being a very useful in many applications such as RICH, TPC detectors, calorimetry, neutron position-sensitive detectors, etc.

## References

1. R.Oliveira, V.Peskov et al., Nucl. Instr. and Meth. **A576** (2007), 362-366.